# Secondary structure prediction of protein constructs using random incremental truncation and vacuum-ultraviolet CD spectroscopy

Mária Pukáncsik[1], Ágnes Orbán[2], Koichi Matsuo[3], Kunihiko Gekko[3], Darren Hart[4] Kézsmárki István[2] and Beáta G. Vértessy [1,5]

[1]Institute of Enzymology, Biological Research Center, Hungarian Academy of Sciences, Budapest, Hungary and [2]Department of Physics, Budapest University of Technology and Economics, Budapest, Hungary, [3] Hiroshima Synchrotron Radiation Center, Hiroshima University, Higashi-Hiroshima, Japan, [4] European Molecular Biology Laboratory, Grenoble, France, [5]Department of Applied Biotechnology, Budapest University of Technology and Economics, Budapest, Hungary

**Abstract**

A novel uracil-DNA degrading protein factor (termed UDE) was identified in *Drosophila melanogaster* with no significant structural and functional homology to other uracil-DNA binding or processing factors. Determination of the 3D structure of UDE will be a true breakthrough in description of the molecular mechanism of action of UDE catalysis, as well as in general uracil-recognition and nuclease action. The revolutionary ESPRIT technology was applied to the novel protein UDE to overcome problems in identifying soluble expressing constructs given the absence of precise information on domain content and arrangement. Nine specimen from the created numerous truncated constructs of UDE were choosen to dechiper structural and functional relationships. VUVCD with neural network was performed to define the secondary structure content and location of UDE and its truncated variants. The quantitative analysis demonstrated exclusive α-helical content for the full-length protein, which is preserved in the truncated constructs. Partition of α-helical boundles comparison with the truncated protein segments denoted new domain boundaries which differ from the conserved motifs determined by sequence-based alignment of UDE homologues in pupating insects. Here we demonstrate that combination of ESPRIT and VUVCD with NN provided significant structural description of UDE and resulted useful truncated constructs for further detailed functional studies.

**Introduction**

Detailed knowledge of protein three-dimensional structure is indispensable for understanding the mechanism of protein action. As of present, macromolecular X-ray crystallography and multidimensional NMR are the techniques of choice to achive this goal. Despite numerous advances in both these methodologies during the recent years, several limitations still exist. For a detailed 3D structural determination by multidimensional NMR, the size of the protein is an important factor, and proteins larger than 30 kDa pose serious difficulaties preventing structural determination **(1)**. There is no such size limitation in macromolecular X-ray crystallography, in this case, however, the need for well-diffracting crystal specimens is still a bothersome bottleneck. Methods to predict and help the design of crystallizable protein constructs are therefore highly required. Generation and investigation of such deletion construcst may then be a first and important step towards characterization of full length proteins.

In the case of multidomain proteins, the easiest way to create shorter constructs is truncation via PCR cloning. In contrast to PCR-based methods that rely on rational design of primers (implying some knowledge of domain structure a priori), the random screening method **E**xpression of **S**oluble **P**rotein by **R**andom **I**ncremental **T**runcation (ESPRIT) uses an exonuclease protocol **(2,3)** to generate all possible gene deletions. Thus every point in the gene is tested as a start or stop position. Large deletions may result in isolated domains, whilst short deletions have been shown to rescue soluble expression of nearly full-length protein **(4,5).** To decide on the well-folded character of the generated truncated constructs, several biophysical methods can be used and the results will then identify potentially promising candidates construcst for further detailed investigations.

In contrast to the above mentioned methods of X-ray crystallography and NMR for the reconstruction of 3D protein structure with atomic resolution, circular dichroism (CD) spectroscopy is generally thought to provide only limited structural information via excitations of peptide bonds by ultraviolet light. CD is mostly used to measure the ratio of the different secondary structure components in proteins **(6-8).** However, recent extensions of CD spectroscopy to the infrared and X-ray photon-energy regions opened new perspectives – by probing molecular vibrations **(9)** and core electron excitations **(10)**, respectively – with the hope of a spatially specific structural analysis. Still, determination of the full secondary or tertiary structure of proteins requires the analysis of the CD spectra using ab initio or neural network calculations based on their amino-acid sequence **(11).**

Our aim in the present study was to probe if a combination of ultraviolet CD spectroscopy with a state-of-the-art biochemical truncating method is capable to map the full secondary structure of proteins solely on an experimental basis without former knowledge about the amino-acid sequence.

Towards this aim, we selected a protein identified as a uracil-DNA degrading factor (UDE) for our approach to combine ESPRIT and VUVCD. UDE was identified in *Drosophila melanogaster*, which recognizes and removes uracil from DNA at the end of the third larval stage **(12)**. The fruitfly genome lacks the otherwise common uracil-DNA glycosylase and, as shown recently **(13)**, can tolerate uracil incorporation in its DNA. We showed that UDE has close homologues only in the genomes of other pupating insects: no function has yet been attributed to any of these homologues. Based on sequence homology, it seems that UDE does not show any similarities to other uracil-processing and -excising glycosylases or nucleases described so far. Therefore, characterization of the structure of this novel DNA-degrading protein is expected to provide significant novel insights into uracil-DNA recognition. Limited proteolysis showed extensive protection by DNA along duplicated conserved sequence Motifs 1A and 1B, and de novo modelling of Motif 1A and 1B predicted similar α-helical bundles and two conserved positively charged surface patches for both motifs. These results suggested that DNA binding may occur at the N-terminal segment of UDE that contains Motifs 1A and 1B **(14).** Despite numerous crystallization efforts, no protein crystal specimens could be generated, perhaps due to the predicted very high degree of conformational freedom of several protein segments **(15).** UDE therefore presented an interesting object of present study.

In the present work, we generated tens of thousands of randomly truncated UDE constructs. Following selection of key best hits, we performed melting and gel filtration studies to evaluate the folded character of the constructs. We analyzed the secondary structural organitazion of these constructs in a comparions with full-length UDE by VUVCD. Quantitative analysis of CD data by SELCON3 program and neural network enabled us to designate new potentially domain boundaries.

# Results

## Construction of ude libraries by ESPRIT

Limited proteolysis data strongly suggested that UDE has a multidomain structure. Since UDE has few useful homologues for multiple sequence alignment construction, we have been unable to accurately define these domain boundaries. Expression of soluble proteins by random incremental truncation (ESPRIT) was used to create unidirectional truncation of full-length UDE **(2-5).**

The sequence-based structural disorder predictions suggested flexible segments at the N-, and C-terminal end of the protein, as well as in the linker region between Motifs 1A and 1B (14). Based on these data we designed six starting point for the unidirectional gene truncation from either the 5' or 3' end of the UDE coding sequence (Figure 1). The six UDE inserts were cloned into the plasmid vector pYUM6002 that locates restriction sites with exonuclease III sensitive and resistant overhangs at the terminus to be truncated and encoding a C-terminal biotin acceptor peptide (BAP) **(16-18)** for indication of soluble expression (Figure 1). Unidirectional gene deletions were achieved by incubation of linearised plasmid with exonuclease III with samples removed every 30 seconds to obtain incremental truncations of ude inserts with a linear size distribution **(17)**. Mung bean nuclease was then added to remove the remaining single-strand and a ligation reaction resulted in fusion of the truncated UDE fragments to a hexahistidine tag with TEV protease cleaving site. Fragments from the six exonuclase III truncation mix were separated by size on agarose gel and divided four (small/large and N-terminal/C-terminal) sublibraries (Figure 2A). Size distribution was verified for each sublibraries (Figure 2B) by amplifying 48 colonies from each libraries with vector specific flanking primers. Truncated constructs of these sublibraries were transformed to *E.coli* BL21 cells for screening expression level and solubility.

## Screening of libraries and scale-up

We sampled every construct 2 times (oversampling) in order to identify truncated constructs exhibiting soluble expression of domains. In the experiment 28,000 colonies were routinely picked into 384-well microtiter plates by a colony-picking robot. Since it is impossible to directly measure soluble expression of thousands of constructs, the method uses a short C-terminal biotin acceptor peptide as a proxy for soluble expression: if the resulting protein fragment is soluble, the endogenous BirA enzyme of *E. coli* mediates efficient in vivo biotinylation **(18).** UDE clones were arrayed by robots onto nitrocellulose membranes for growth of colony arrays. Identification of hits was by hybridisation of the arabinose-induced colony blot with a fluorescent streptavidin conjugate. Batches of 96 hits (Figure 2B) from this analysis were expressed in 4 ml cultures in 24-well plates and purified on $Ni^{2+}$ NTA resin using a Tecan liquid handling robot. Eluted fractions were then analysed by high throughput SDS-PAGE to confirm solubility and purifiability (Figure 2C). The 72 positive clones were sent then to be sequenced to identify the soluble regions of the protein.

Based on the amino acids alignment of the sequenced truncated UDE clones nine cluster of constructs with similar size were identified. Compared to the previously designated conserved motifs determined by the alignment of UDE homologues sequences we identified new domain boundaries- even if with only a few amino acids difference -in the case of five fragments (Figure 3A). The expression level in small scale and the covered UDE conserved segments were taken into consideration when one truncated UDE fragments of each cluster were choosen for scale up and further structural analysis. After optimization the expression conditions the nine members of each clusters in large scale (500 ml) we got well-expressing and purifying truncated proteins (Figure 3B). To check that the proteins are not aggregated and to analyse their oligomer status we used analytical gelfiltration. We observed that the proteins formed dimer in solution except the NC6 fragment, that forms a monomer (Table 1.). In the case of the CA7 construct soluble aggregation was observed, because a single peak in the exclusion volume of the column was detected during elution. We also performed detailed analysis of UDE deletion mutants on thermostability by ThermoFluor. Determination the melting temperature of the UDE fragments gave partial information on protein folding. CA7 and CH9 constructs did not show characteristic meltingpoint suggesting these fragments does not have well-defined tertiary/quaternary structure (Table 1.). It is also confirmed in the gelfiltration experiment, where behaviour of CA7 fragment diverged from typical for globular structures and it probably contains unfolded, flexible segments.

**Secondary structure estimation by combining VUVCD results with neural network (NN) algorithm**

For detailed secondary structural studies vacuum-ultraviolet circular dichroism (VUVCD) spectroscopy measurements were carried out on the full length UDE and its nine truncated fragments over the wavelength region of 160-260 nm using the bright synchrotron-radiation of the Hiroshima Synchrotron Radiation Center **(19)**. Such extension of the conventional far-ultraviolet CD spectroscopy towards shorter wavelengths highly improves the reliability of the estimation of the protein secondary structure content **(20,21)**.

The spectral shape for UDE, shown in Figure 4A, indicates high α-helix content. Similarly to the CD spectra of regular α-helices, the spectrum of UDE is characterized by two negative peaks at 222 and 208 nm, and a positive peak at 192 nm **(22)**. The minimum at 222 nm is due to the nπ* transition of the carbonyl group of peptide, while the parallel and perpendicular excitation of the peptide ππ* transition is responsible for the other minimum at 208 nm and the positive peak at 192 nm **(8,23)**. Additional negative band at around 170 nm and a positive shoulder at 175 nm were found specific to α-helical structures by former VUVCD measurements **(20,21)**. Though these characteristics imply exclusive α-helix content for the intact protein, the intensity of the CD spectrum is reduced by about a factor of two as compared with purely α-helical structures, which suggests that a considerable portion of the sequence forms different secondary structures or remains unordered.

Quantitative evaluation of the spectral data was performed by the SELCON3 program with the VUVCD spectra of 31 reference proteins over the wavelength region of 160-260 nm **(24,25).** We found the following secondary structure contents: 62% α-helix (40% regular and 22% distorted helix), 8% β-strand (2% regular and 6% distorted strand), 9% turn and 21% unordered component as shown in Figure 4B. From the portion of the distorted parts located at the edges of the regular regions, we estimate the number of α-helical segments in UDE to be $n_\alpha=20$ **(26,27).** On the other hand, the anomalously high portion of distorted parts in β-strands may imply that the corresponding structures cannot be unambiguously classified as β-strands and are hardly distinguished from unordered structures. We have performed a neural network analysis treating the amino-acid sequence and the secondary structure contents (obtained from the CD spectrum) on equal ground to determine the location of the α-helical and β-strand segments, i.e. to obtain the spatially resolved secondary structure of UDE **(28).** The resulting structure, schematically displayed in Figure 5A-B, reflects good correspondence between the CD spectrum and the amino-acid sequence, since the portion of each secondary-structure component has been affected by less than 2-3%. Though the presence of short β-strand regions is supported by the neural network algorithm, we found that this method strongly underestimates the number of α-helical and β-strand segments in comparison with the results solely based on the CD spectrum of the protein. Such tendency to merge neighbouring α-helical or β-strand segments separated by a few amino acids with unordered structure have already been reported in the literature **(28).**

As discerned in Figure 4A, the CD spectra of the nine fragments of UDE indicate that the exclusive α-helical content is preserved in the fragments, especially in case of NA1, NC6, NG3, NA3. Indeed the results of the neural network analysis, shown in Figure 5A-B, verifies that the native structure is conserved in the inner regions of the fragments, except for CA7 where the truncation process is followed by a drastic transformation of the structure. By averaging the results obtained for the fragments with those of the full-length protein in the overlapping regions, we could further improve the reliability of the secondary structure estimation for UDE. Our final estimate for the secondary structure of UDE is given in Figure 5C. From the averaging process the edge regions of the fragments (five amino acids) and the full CA7 were excluded, since they do not preserve their native structure.

Taking advantage of the structural stability characteristic to most of the UDE fragments, we demonstrate that the secondary structure can be spatially resolved along the sequence solely based on the CD spectra. As an example, we investigate the native structure of the N- terminal end of the protein (see Figure 5D), since the neural network analysis predicted significant difference between this part in the intact protein and the corresponding CA7 fragment. We estimate the CD spectrum of the N- terminal as the difference between the CD spectra of UDE and NA2 according to $(\Delta\varepsilon_{UDE} \times N_{UDE} - \Delta\varepsilon_{NA2} \times N_{NA2})/(N_{UDE} - N_{NA2})$, where $\Delta\varepsilon$ is the molar ellipticity and N is the number of amino acids for UDE and NA2. This difference spectrum, displayed in Figure 4B, clearly shows that the N-terminal is purely α-helical (74% regular helix, 18% distorted helix, 4% regular strand and 4% unordered) when embedded in the full-length protein. The efficiency of this subtraction method is more obvious when comparing highly overlapping fragments such as NG3 and NA3 as indicated in Figure 5D.

The difference spectrum corresponding to an only 25-amino-acid-long part of the sequence again shows exclusive α-helix content (65% regular helix, 15% distorted helix and 20% turn). Besides the high quality CD spectral information, we emphasize that the conservation of the native structure is required for the fragments used in such comparisons.

**Discussion**

In this study we partially continued the structural characterization of the unique uracil-DNA degrading factor utilizing the preliminary domain organization analysis and secondary structure prediction for the N-terminal end of UDE.

Secondary structure prediction suggested that the duplicated fragments (Motifs 1A-1B) are mainly alpha-helical and interact through a conserved surface segment. Structural disorder predictors classified that the UDE protein possesses flexible segments at both the N- and C-termini, and also in the linker regions of the conserved motifs. However, the secondary structural content of the well-folded C-terminal domain (containing Motifs 2-4) was still not clear **(14)**.

The ESPRIT-method based library screening **(3)** was applied to UDE and we obtained well-folded, highly expressing, soluble truncated UDE fragments. We discerned nine protein cluster and the protein folding, thermostability and oligomer status of the nine proteins represent certain cluster were examined. The analysis indicated that CH9, CH10, CA7 variants though the well expression characteristics contain disordered or flexible residues that potentially inhibit well-defined tertiary/quaternery structure formation.

Using multiple seqence alignments we found that five clusters determine potential new domain boundaries compared to the distribution of conserved motifs in UDE homologs from pupating insects. The CH9 and CH10 constructs comprising the previously unpredicted domain termination with different terminal residue would inhibit the chances of crystallization and functional characterization based on the result of biophysical investigation.

We performed detailed analysis of UDE deletion mutants on secondery structure content used the data of vacuum-ultraviolet circular dichroism (VUVCD) spectroscopy **(19-21)**. Quantitative evaluation of VUVCD data by SELCON 3 **(24-27)** and neural network algorithm **(28)** determined the number and location of α-helical segments in the fragments and total length UDE. Data confirm the mainly α-helical content for the full-length protein, which is preserved in the truncated constructs, also for the compact C-terminal end of UDE. Comparison of secondary structure partition of the N-terminal duplicated domains it is evident that the previously determined boundaries of conserved motifs do not correspond exactly to the predicted arrangement of α-helical segments.

α-helical structure has a significant role in DNA binding motifs such as helix-turn-helix **(29)**, leucine zipper/coiled-coil **(30)** and zinc finger motifs **(31)**. Former results showed that both the total length UDE and its truncated physiologically occurring UDE isoform which was generated from the recombinant UDE by cloning can bind normal and uracil-containing DNA, but degrades only uracil containing DNA **(12,14)**. We suppose that the helix-turn-helix nucleic acid binding structural motif may be present in the full- length UDE and in its constructs processed by random incremental truncation, which is formed either by the positively charged N-terminal segments or the α-helical C-terminal domain. It is confirmed by the oligomerization of the truncated constructs, that may represent the tertiary helix-helix interactions in the full-length protein.

We showed that a combination of ESPRIT and VUVCD provided important insights into the structural description of UDE, used in our study as a model protein with numerous unstructured segments and at least partially independently folding domains. A major advantage of VUVCD is that it may be applicable to dilute solutions of proteins with large molar mass, where X-ray and nuclear magnetic resonance techniques fail.

## Materials and methods

### Cloning of UDE inserts into pYUM6002 vector

Three starting constructs were generated by amplifying the ude gene with 1 forward primer (UDEfor1 gatcctagggcgcgccgatgattaagtgccatatgccgtcgagttggagacggc) and 3 reverse primers (UDErev1:gatcctagatgcattctccctcttcttcttccttttgggc,UDErev2:gatcctagatgcattcagcttttcgatgtactgcttcagc, UDErev3: gatcctagatgcattgaggaaatcctcaagtacctggactcc), the latter being in frame with the biotin tag (solubility detection) and encoding 3 separate C termini.PCR products were cloned into pYU6002 with AscI and NscI restriction enzymes for N-terminal truncation (5' DNA deletion) library construction. Similarly, another three starting constructs were generated by amplifying the ude gene with 1 reverse primer (UDErev4: gatcctaggcggccgctcactcctccctcttcttcttcctttg) and 3 forward primers (UDEfor2: gatcctaggacgtcgatgattaagtgccatatgccgtcgagttggagacg,UDEfor3:gatcctaggacgtcgtggagacggctacgcaaaatcagt,UDEfor4: gatcctaggacgtcgaatggcggaggggcgtccagc), the latter being in frame with the histidine tag and encoding 3 separate N termini. PCR products were cloned into pYUM6002 with AatII and NotI restriction enzymes for C-terminal truncation (3' DNA deletion) library construction.

### UDE truncation library synthesis

High quality, unnicked plasmid were prepared from 200 ml overnight culture with alkaline lysis protocol and phenol chloroform extraction, propanol precipitation. To remove salts and RNA a Qiagen Minprep Kit was used. For N-terminal truncation 10 μg of the three different pYUM6002 plasmids were digested with *Aat*II and *Asc*I and for C-terminal trucation 10 μg of the other three different pYUM6002 plasmids were digested with *Nsi*I and *Not*I, yielding an exonuclease III insensitive and an exonuclease sensitive end. 4μg linearised plasmid DNA (final conc 33.3 ng/μl) were incubated in NEB buffer 1, 70 mM NaCl and 400 U of Exonuclease III (at 100 U/μl) in a final volume of 120 μl at 22°C PCR machine **(32,33).** To ensure even fragment distribution, every 1 minute, 1/60th reaction volume (2 ul) was transfered to a quenching tube (200 μl of 3 M NaCl) on ice. The quenched reaction was denatured at 70°C for 20 min. UDE fragments were purified using Machery-Nagel Nucleospin Extract II kit according to their protocol. In order to remove the 5′ overhang after exonuclease III digestion, the purified UDE fragments were treated with 5 U of Mung Bean Nuclease in 1XMung Bean buffer at 30°C for 30 min, then cleaned up with the Nucleospin Extract II kit. The ends of the UDE fragments were polished by incubation with 5U Pfu polymerase in 1×Pfu polymerase native buffer, 2.5 mM dNTPs) at 72°C for 20 min. For UDE-Library size fractionation the N-terminal and C-terminal reactions were electrophoresed in a 0.5% agarose gel and inserts with defined size were excised. We divided each library in 2 sub-libraries: big size and small size. UDE fragments were extracted using QIAexII Kit and recircularised with T4 DNA Ligation Kit. Constructs of the four sub-libraries were transformed OMNImax competent cells. Transformation mixes were recovered in SOC medium and plated on LB agar Petri dishes supplemented with kanamycin 50 μg/ml. Remainder of SOC mix was placed at 4 ºC.

**Colony blot analysis**

*E. coli* BL21 (DE3) competent cells were transformed with the four UDE sub-libraries (32,33). The sub-libraries were plated in 22 cm LB agar QTrays (supplemented with kanamycin and chloramphenicol) at a density of 2000, 4000 and 8000 colonies per tray. The 12 QTrays were then incubated overnight at 37°C. From both N-terminal and C-terminal small and big libraries approximately 28 000 colonies were isolated using a colony-picking robot into 384 well plates containing 80 µl TB medium per well (supplemented with kanamycin and chloramphenicol). Liquid cultures were grown overnight in a HiGro at 37°C with 300 rpm of shaking. All clones were gridded robotically onto a nitrocellulose membrane on LB agar plates (supplemented with kanamycin and chloramphenicol). The gridded membranes were incubated overnight at 25°C until colonies on the membrane were just visible. Next day the membranes were moved onto a fresh pre warmed LB agar plate (supplemented with antibiotics, arabinose 0,2% final concentration and 50 µM biotin) to induce recombinant protein expression at 30°C for 4-5 h avoiding to let colonies merge. Membranes were placed on filter paper soaked in denaturing buffer (0.5 M NaOH, 1.5 M NaCl) and incubated 10 minutes at room temperature to lyse colonies. The membranes were neutralised for 5 min in neutralization buffer (1 M Tris, 1,5 M NaCl, pH 7.5) and immersed in 2×SSC buffer for 15 min. The remaining colony debris on the membranes were carefully removed with a glass spreader. The membranes were blocked overnight in Superblock at 4°C in Hibridyser. The membranes were washed with PBS-Tween (PBS with 0.1% Tween 20) buffer for 3X15 minutes and incubated in 50 ml PBS-Tween containing 16 µl anti-hexahistidine antibody for 1 h at 4°C. After washing steps with PBS-Tween buffer, the membranes were incubated in 50 ml of PBS-Tween containing 10 µl streptavidin Alexa Fluor 488 and 50 µl Alexa Fluor 532 rabbit anti-mouse IgG for 1h at 4°C. After 3X15 minutes washing with PBS-Tween buffer and 5 minutes with destilled water, the membranes were scanned with a Typhoon 9400 fluorescence scanner for hexahistidine tag and biotin acceptor peptide signal intensities respectively. Signals were quantified from digitised images using Visual Grid software and data exported to Microsoft Excel for analysis.

**High-throughput protein expression and purification**

The best 96 clones from the UDE libraries were selected for small-scale protein expression in 4 ml LB in 24 well plates. Protein expression was induced at OD600 = 0.7 with arabinose 0,2% final concentration and 50 µM biotin overnight at 25°C. Next day the cells were pelleted by centrifugation and resuspended in 4 ml of spheroplast buffer (20 mM Tris, 250 mM NaCl, 20% sucrose and 1 mg/ml lysozyme, pH 8.0). The sphaeroplasted cells were centrifuged at 3700 rpm for 10 minutes at 4°C and then resuspended in 800 µl lysis buffer ( 10 mM Tris, 0.5% Brij, 0.25 U/µl Benzonase, pH 7.5and 0.8 µl Protease Inhibitor Cocktail). Lysated protein samples were simultaneously purified by Tecan liquid handling robot. Protein samples were loaded onto a 96 well filter plate supplemented with 50 µl Ni2+NTA resin in each well and mixed by rotation at 4°C for 30 min. The samples were washed with 50 mM Sodium phosphate buffer, pH 7, 300 mM NaCl, 5 mM imidazole and eluted with 50 mM Sodium phosphate buffer, pH 7, 300 mM NaCl, 300 mM imidazole. The purified 96 samples were visualised on High-Throughput SDS–PAGE and their sequence boundaries were determined by DNA sequencing **(32,33).**

**Scale-up expression and purification of selected clones**

From the 72 positive clones 9 clones were scaled up using the *BL21(DE3)ung-151pLysS E. coli* expression system in 500 ml LB at 37°C. Protein expression was induced at OD600 = 0.7 with 0,1 mM IPTG at 25°C for 4-5 h. The recombinant His-tagged UDE proteins were purified on a HisTrap column in an AKTA Purifier System, with unicorn software. Absorbance of eluates was monitored at both 260 and 280 nm. Total cell lysate was prepared from 500 mL of culture in 25 mL of lysis buffer (50 mM Tris∕HCl, 150 mM KCl, and 1 mM EDTA, pH 8.0, complemented with 1 mm dithiothreitol, 0.1 mm phenylmethanesulfonyl fluoride and proteinase inhibitor cocktail). After loading, the HisTrap column was washed with buffer A (50 mM Tris∕HCl, 30 mM KCl, 1 mM EDTA, and 5 mM imidazole, pH 7.5, complemented with 0.1 mm phenylmethanesulfonyl fluoride and 1 mm dithiothreitol) until and then with 50% buffer B (buffer A also containing 1 M KCl) to elute the contaminating proteins; finally, a linear gradient of buffer C (buffer A also containing 1M imidazole) was applied. Protein fractions were dialyzed against 25 mM Tris∕HCl, 150 mM KCl, and 1 mM EDTA (pH 7.5), concentrated on a Vivaspin centrifugal concentrator.

**Analytical gel filtration**

Gel filtration was conducted on Superdex 200HR size exclusion column mounted in an AKTA Purifier System, with unicorn software. Absorbance of eluates was monitored at both 260 and 280 nm. Size exclusion chromatography was performed as described previously **(14).**

**Thermofluor**

A thermofluor assay was used to monitor the thermal stability of purified UDE truncated constructs. The reaction mix contained 0,25 mg/ml or 0,5 mg/ml or1 mg/ml of purified UDE fragments, 1 μl of 40x diluted SyproOrange dye, and 25 mM Tris∕HCl, 150 mM KCl, and 1 mM EDTA (pH 7.5) in 25 μl final volume. The mix was put into a 96 well plate and analysed with a Stratagene Mx3000P real-time PCR Thermal Cycler operated by mxpro software. This system contains a heating device for accurate temperature control set at 25°C to 99$^o$C at a ramp rate of 0,5°C /min.

**Vacuum ultraviolet circular dichroism (VUVCD) spectroscopy**

Investigation of the secondary structure of UDE by means of VUVCD spectroscopy has been performed at beamline #15 of Hiroshima Synchrotron Radiation Center (HiSOR) using a high photon flux of 1010 photon/second. Circular dichroism spectra were measured using a polarization modulation technique over the wavelength range of 160-260 nm with a resolution of 1 nm. Details of the setup are reported elsewhere **(19).** Due to rapidly increasing water absorption for wavelength shorter than 180 nm, we kept the light path within the samples as short as ~12 μm and used solutions in 10 mM NaPi (pH=7.5) buffer with the protein concentrations 6 mg/ml or 9 mg/ml and 3 mg/ml in the case of CA7.

**Secondary structure analysis**

The secondary structure components and the number of segments of full length UDE and UDE constructs were determined using the SELCON3 program and the VUVCD spectra of 31 reference proteins with known X-ray structures **(24-27).** The sequence based prediction of the positions of the secondary structuce components along the amino acid sequences of UDE and UDE constructs were carried out by combining VUVCD analysis with the NN algorithm. The VUVCD-NN method was detailed previously **(28).**

# Figures

**Table 1**

| UDE constructs | $T_m$ (°C) | Analytical gelfiltration | | | Oligomer status |
|---|---|---|---|---|---|
| | | $V_{el}$ (kDa) | $M_{app}$ (kDa) | $M_{calc}$ (kDa) | |
| NA1 | 51,3 | 62,4 | 42,1 | 25,7 | dimer |
| NA2 | 53,9 | 56,6 | 57 | 26,3 | dimer |
| NA3 | 57,1 | 69,6 | 28,9 | 18 | dimer |
| NC6 | 53,7 | 55,7 | 35,6 | 34,7 | monomer |
| NG1 | 56,4 | 60 | 47,7 | 23,3 | dimer |
| NG3 | 51,5 | 64,6 | 37,5 | 20,9 | dimer |
| CA7 | - | 44,3 | - | 25,4 | monomer |
| CH9 | - | 58,3 | 52,1 | 22,1 | dimer |
| CH10 | - | 56 | 58,8 | 32.1 | dimer |
| full length UDE | | | | | monomer |

**Figure 1**

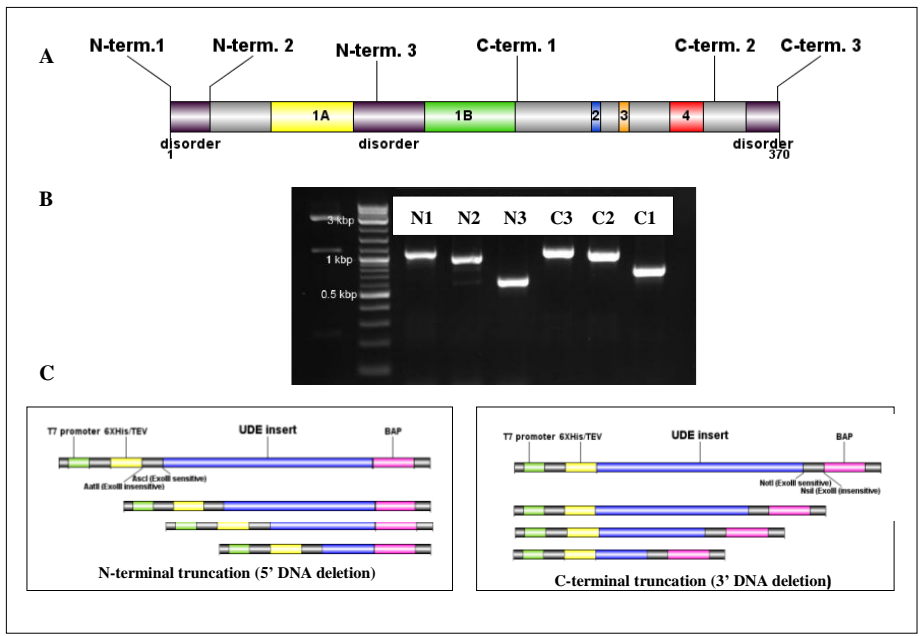

Figure 2

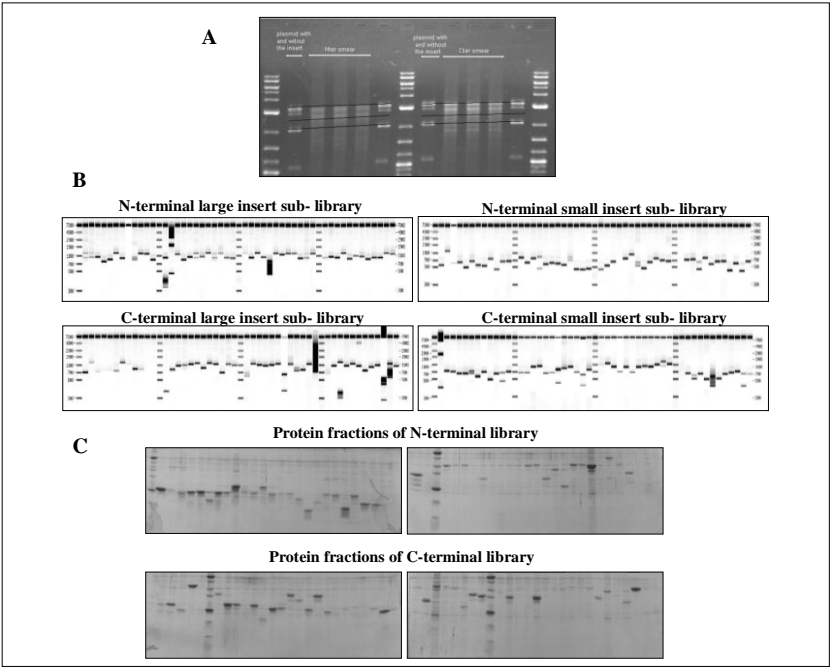

Figure 3

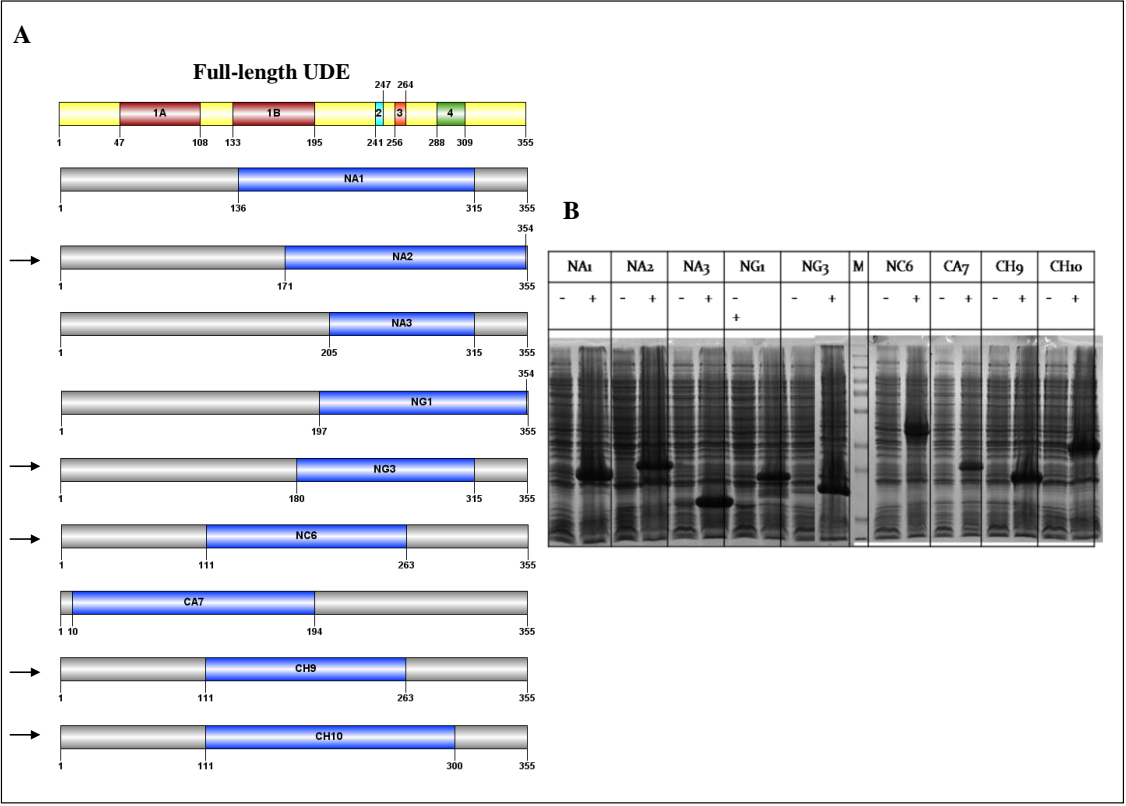

**Figure 4**

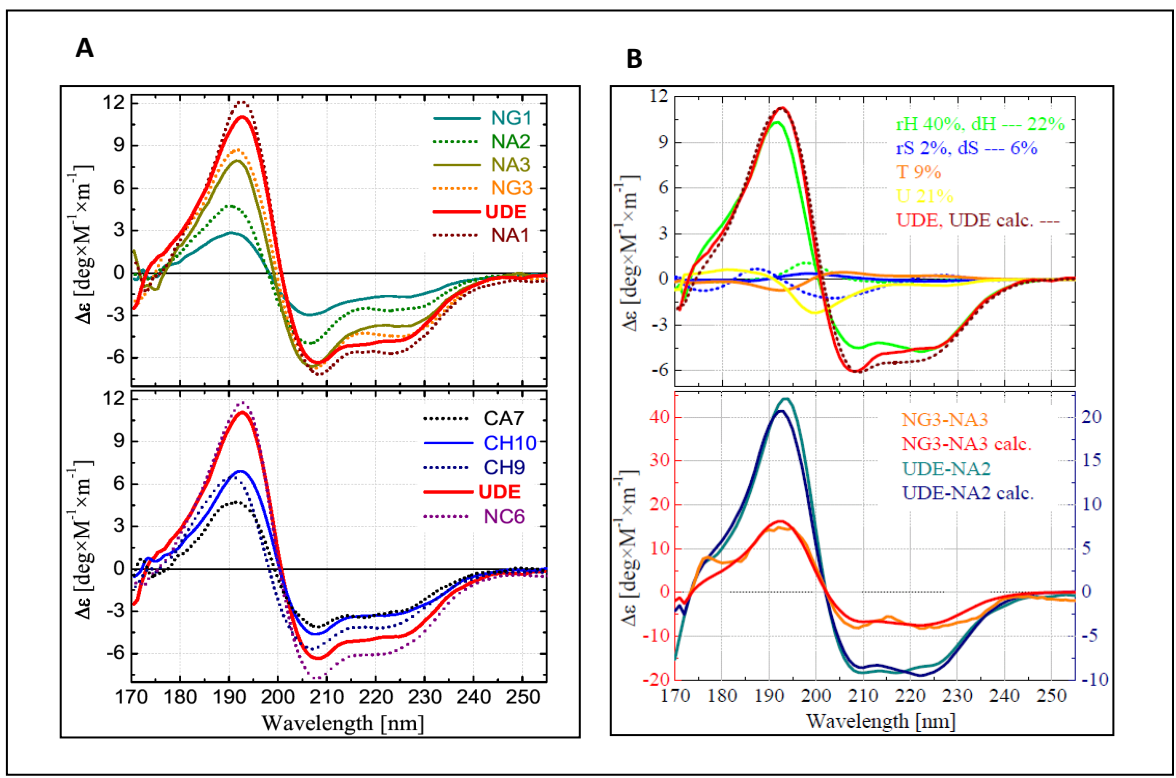

**Figure 5**

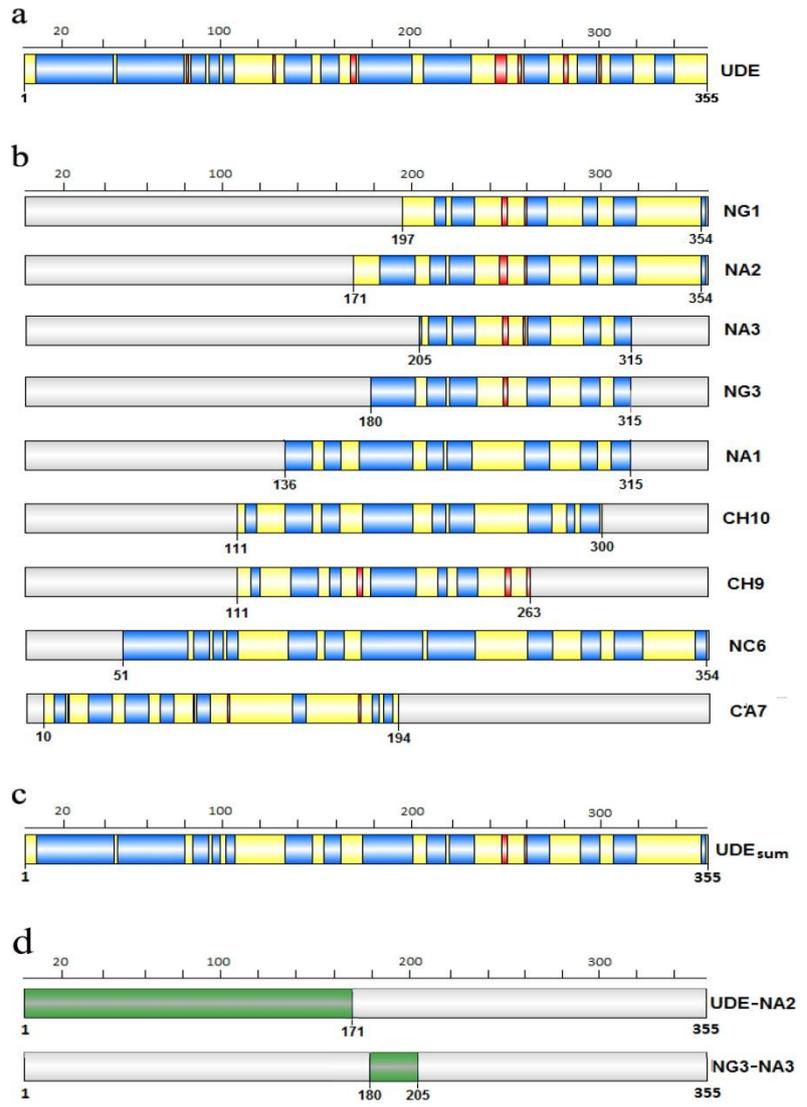

**Tables**

**Table 1.**

The results of analytical gelfiltration and the determination of the oligomer status of each truncated fragments. Vel: elution volume, Mapp: apparent molecular weight, Mcalc: calculated molecular weight based on column calibration.

Data of thermostability analysis by Thermofluor. Tm: melting point of the given protein.

**Figure legends**

**Figure 1: (A)** The location of designed starting points along the UDE sequence indicating the predicted disordered segments and the conserved motifs. **(B)** The corresponding UDE inserts amplified in PCR reaction. **(C)** The truncated UDE gene fragments generated by N-terminal truncation were fused in - frame with the biotin acceptor peptide and out-of-frame with hexahistidine tag, while fragments produced by C-terminal truncation were fused in-frame with hexahistidine tag and out-of-frame with BAP.

**Figure 2: (A)** Size fractionation of ude fragments generated by unidirectional truncation on agarose gel. In the lanes next to the DNA ladders is the vector with total length ude gene at higher position while the empty vector is at a lower position. N1-N3 and C1-C3 marked samples show by the exonuclease III. truncation generated ude constructs. **(B)** Assessment of ude sublibraries size and diversity by PCR screen. **(C)** Separation of purified protein fractions on Ni2+-NTA resin from N-terminal (upper panels) and C-terminal (bottom panels) libraries on HTP SDS-PAGE.

**Figure 3: (A)** The restricted nine UDE truncated fragments from the identified protein clusters that was choosen for scale-up. Arrows show the detected new domain boundaries compared to the previously designated conserved motifs determined by the alignment of UDE homologues sequences. **(B)** The optimized expression of the nine UDE constructs in E.coli BL21 cells before (-) and after (+) IPTG induction.

**Figure 4: (A)** Vacuum-ultraviolet circular dichroism (Δε) spectra of the UDE protein and its nine truncated fragments measured over the wavelength region of λ=160-260 nm. The spectra are sorted into two panels for better visibility and the data for UDE are displayed in the both for reference.

**(B)** Decomposition of the CD spectra of UDE and its selected fragments using six secondary structure components; regular/distorted α-helix (rH/dH), regular/distorted β-strand (rS/dS), turn (T), and unordered structure (U). Upper panel: CD spectrum of UDE as measured and as fitted using the six components **(24-26).** Spectra of the components are also plotted with magnitudes proportional to their ratios in the full protein. Lower Panel: Difference spectra corresponding to UDE-NA2 and NG3-NA3 together with the fittings based on the spectra of the six basic components.

**Figure 5:** Spatial distribution of the secondary structure components in **(A)** the full-length (355 amino acids) UDE protein and **(B)** in its nine truncated fragments as determined from the CD spectra and the amino-acid sequence using a neural network algorithm **(27-28)**. α-helical segments and β-strands are displayed in blue and red, respectively, while both turns and unordered parts appear in yellow. **(C)** Our final estimate for the secondary structure of UDE obtained as an average of the structure of UDE proposed in panel **(A)** and the structure of the fragments shown in panel **(B)** except for CA7. **(D)** The native structure of the N- terminal end of the full length UDE was investigated using the evaluation of the CD spectrum of the N-terminal as the difference between the CD spectra of UDE and NA2 according to $(\Delta\varepsilon_1 UDE \times N_1 UDE - \Delta\varepsilon_1 NA2 \times N_1 NA2)/(N_1 UDE - N_1 NA2)$, where $\Delta\varepsilon$ is the molar ellipticity and N is the number of amino acids for UDE and NA2. The same subtraction method was performed with the highly overlapping fragments NG3 and NA3.